\documentclass[aps,prx,reprint,superscriptaddress]{revtex4-2}

\usepackage{graphicx}
\usepackage{amsmath}
\usepackage{braket}
\usepackage{xcolor}
\usepackage[normalem]{ulem}
\usepackage{notes2bib}
\bibnotesetup{
note-name = ,
use-sort-key = false
}
\usepackage[nameinlink,capitalise]{cleveref}
\newcommand{\expect}[1]{\langle #1 \rangle}

\begin{document}

\title{Metal-insulator transition and magnetism of SU(3) fermions in the square lattice}

\author{Eduardo Ibarra-Garc\'ia-Padilla}
\email[]{edibarra@ucdavis.edu}
\affiliation{Department of Physics, University of California, Davis, CA 95616, USA}
\affiliation{Department of Physics and Astronomy, San Jos\'e State University, San Jos\'e, CA 95192, USA}
\affiliation{Department of Physics and Astronomy, Rice University, Houston, Texas 77005-1892, USA}
\affiliation{Rice Center for Quantum Materials, Rice University, Houston, Texas 77005-1892, USA}
\author{Chunhan Feng}
\affiliation{Center for Computational Quantum Physics, Flatiron Institute, 162 5th Avenue, New York, NY 10010, USA}
\author{Giulio Pasqualetti}
\author{Simon F\"olling}
\affiliation{Ludwig-Maximilians-Universit{\"a}t, Schellingstra{\ss}e 4, 80799 M{\"u}nchen, Germany}
\affiliation{Max-Planck-Institut f{\"u}r Quantenoptik, Hans-Kopfermann-Stra{\ss}e 1, 85748 Garching, Germany}
\affiliation{Munich Center for Quantum Science and Technology (MCQST), Schellingstra{\ss}e 4, 80799 M{\"u}nchen, Germany}
\author{Richard T. Scalettar}
\affiliation{Department of Physics, University of California, Davis, CA 95616, USA}
\author{Ehsan Khatami}
\affiliation{Department of Physics and Astronomy, San Jos\'e State University, San Jos\'e, CA 95192, USA}
\author{Kaden R. A. Hazzard}
\affiliation{Department of Physics and Astronomy, Rice University, Houston, Texas 77005-1892, USA}
\affiliation{Rice Center for Quantum Materials, Rice University, Houston, Texas 77005-1892, USA}
\affiliation{Department of Physics, University of California, Davis, CA 95616, USA}

\date{\today}

\begin{abstract}
We study the SU(3) symmetric Fermi-Hubbard model (FHM) in the square lattice at $1/3$-filling using numerically exact determinant quantum Monte Carlo (DQMC) and numerical linked-cluster expansion (NLCE) techniques. We present the different regimes of the model in the $T-U$ plane, which are characterized by local and short-range correlations, and capture signatures of the metal-insulator transition and magnetic crossovers. These signatures are detected as the temperature scales characterizing the rise of the compressibility, and an interaction-dependent change in the sign of the diagonal spin-spin correlation function. The analysis of the compressibility estimates the location of the metal-insulator quantum critical point at $U_c/t \sim 6$, and provides a temperature scale for observing Mott physics at finite-$T$. Furthermore, from the analysis of the spin-spin correlation function we observe that for $U/t \gtrsim6$ and $T \sim J = 4t^2/U$ there is a development of a short-range two sublattice (2-SL) antiferromagnetic structure, as well as an emerging three sublattice (3-SL) antiferromagnetic structure as the temperature is lowered below $T/J \lesssim 0.57$. This crossover from 2-SL to 3-SL magnetic ordering agrees with Heisenberg limit predictions, and has observable effects on the density of on-site pairs. Finally, we describe how the features of the regimes in the $T$-$U$ plane can be explored with alkaline-earth-like atoms in optical lattices with currently-achieved experimental techniques and temperatures. The results discussed in this manuscript provide a starting point for the exploration of the SU(3) FHM upon doping.
\end{abstract}

\maketitle

\section{Introduction}

The study of Fermi-Hubbard models (FHMs) with spin $S>1/2$ and higher SU($N>2$) symmetries has gained interest in recent years due to the interesting exotic ground states they are predicted to display, their relevance to multi-orbital materials, and their precise realization in ultracold atoms. Their richness is exemplified in the Heisenberg limit, where the density is on average one particle per site ($\expect{n} =1$) and the repulsive on-site interaction energy dominates over the tunneling amplitude ($U \gg t$). Already in this simplifying limit, the model on the two-dimensional (2D) square lattice is predicted to exhibit a plethora of ground states with a complicated $N$-dependence~\cite{Tokura2000,Toth2010,Bauer2012,Hermele2009,Hermele2011,Nataf2014,Corboz2011}.

The possibility to experimentally explore the rich phase diagram of the SU($N$) FHM is opened by the rapid development of quantum simulation capabilities with alkaline-earth-like atoms (AEAs) in optical lattices (OLs). Due to the intrinsic SU($N$) nuclear spin symmetry of fermionic AEAs (such as $^{173}$Yb and $^{87}$Sr)~\cite{Wu2006,Cazalilla2009,Gorshkov2010,Cazalilla2014,Stellmer2014}, experiments can engineer the SU($N$) FHM with $N$ tunable from $2,3,\ldots,10$. In the past few years, experiments with $^{173}$Yb in OLs have explored the SU($N$) FHM for different values of $N$ and in different geometries: In the 3D cubic lattice, they have studied the Mott insulator state for $N=6$~\cite{Taie2012}, the equation of state for $N=3,6$~\cite{Hofrichter2016}, and a flavor-selective Mott insulator for $N=3$~\cite{Tusi2021}. In 2D dimerized OLs, experiments have measured nearest-neighbor antiferromagnetic (AFM) correlations for $N=4$~\cite{Ozawa2018}. In 1D, 2D, and 3D hypercubic lattices with uniform tunneling rates, nearest-neighbor AFM correlations have been detected for $N=6$~\cite{Taie2020}. The future implementation of quantum gas microscopy for two-dimensional OLs realizing the SU($N$) FHM~\cite{Chiu2019,Altman2021,Gross2017,Bloch2012,Parsons2016,Mazurenko2017,Koepsell2020,Hartke2020,Ji2021} is expected to reveal a wealth of physics as it will allow for the characterization of finite-temperature analogs of a variety of proposed ground states via the direct observation of site-resolved spin-spin correlations~\cite{Yamamoto2016,Taie2020,Schafer2020,Okuno2020}. These experimental efforts call for a thorough theoretical understanding of the phase diagrams for $N>2$, and in particular, finite-temperature signatures of their rich physics.

A question of interest in the SU($N$) FHM occurs at $1/N$-filling ($\expect{n}=1$) in the 2D square lattice. At this filling, the $N>2$ FHM provides us with the opportunity to disentangle the role played by nesting and Mott physics since, in contrast to SU(2), a finite $U$ is required to open a charge gap. While the SU(2) FHM at $1/2$-filling exhibits perfect nesting of the Fermi surface and a van Hove singularity in the density of states, the $N>2$ counterparts achieve $\expect{n}=1$ without these special band structure features and allow us to separate band structure attributes from the effect played by interactions without the need to consider next-nearest-neighbor tunneling amplitudes $t'$ which are hard to control in ultracold atom experiments. 

\begin{figure}[tbp!]
\includegraphics[width=0.9\linewidth]{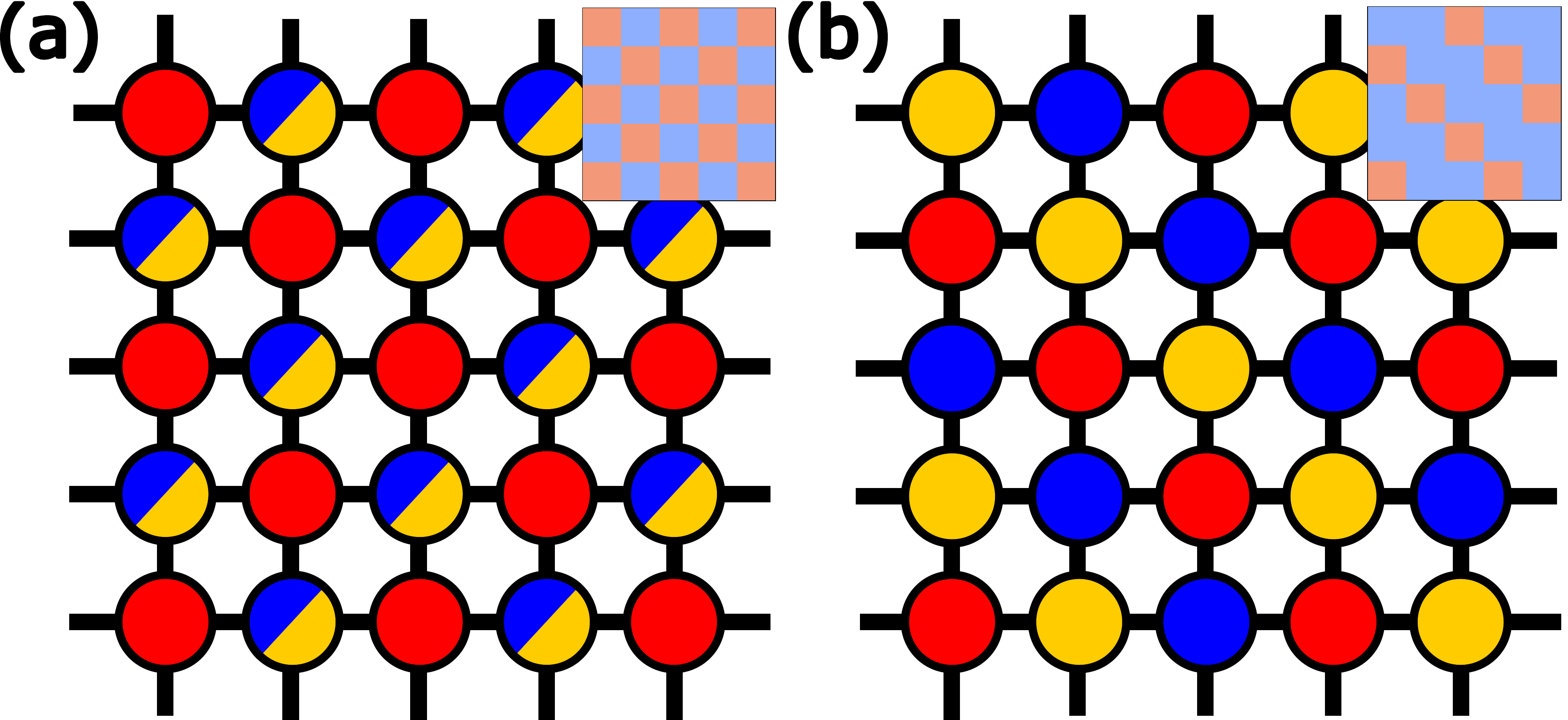}
\caption{\textbf{Schematics of magnetic orderings.} (a) 2-SL AFM. Here the ordering wavevector is $(\pm \pi,\pm \pi)$, where in sublattice $A$ there is a predominance of a spin flavor and in sublattice $B$ there is a spin with a different flavor (a superposition of the other two spin flavors). (b) 3-SL AFM. Here the ordering wavevector is $(\pm 2\pi/3,\pm 2\pi/3)$ and in this case, spins of the same flavor alternate every three sites. The insets correspond to the sign of $C(\vec{r}\,)$ in Eq.~\eqref{eq:sr}, measured from the site in the lower left corner, for the ideal arrangements shown. Orange (blue) corresponds to a positive (negative) sign.}\label{fig::2SL_3SL} 
\end{figure}

In this work, we focus on $N=3$, for which the model is predicted to display interesting physical phenomena even in very special limits and geometries~\cite{Sotnikov2015,Sotnikov2014,Yanatori2016,IbarraGarciaPadilla2021,Nie2017,Hafez2018,Hafez2019,Hafez2020,PerezRomero2021}. For example, in the 2D square lattice in the weak-coupling regime, mean-field theory and renormalization group calculations predict that the ground state is an SU(3) symmetry breaking phase at half-filling ($\expect{n}=1.5$)~\cite{honerkamp2004}. Moreover, at $1/3$-filling ($\expect{n}=1$), our focus in this paper, exact diagonalization, density-matrix renormalization group, and infinite projected entangled-pair states calculations find that the ground state in the Heisenberg limit exhibits a 3-SL AFM ordering~\cite{Toth2010,Bauer2012}. This finding is corroborated by high-temperature series expansions and large-scale exact diagonalization calculations in Ref.~\cite{Romen2020}, which further suggest the existence of a short-range 2-SL AFM ordering prior to the onset of the 3-SL AFM ordering as the temperature is lowered. Furthermore, approximate techniques, such as dynamical mean-field theory, predict such change in the magnetic ordering also occurs in the Hubbard regime~\cite{Sotnikov2014,Sotnikov2015} (see Fig.~\ref{fig::2SL_3SL} for schematics of the magnetic orderings).  

Here, we study the SU(3) FHM on the square lattice at $1/3$-filling, using numerically exact finite-temperature determinant quantum Monte Carlo (DQMC)~\cite{Blankenbecler1981,Sorella1989} and numerical linked-cluster expansion (NLCE) techniques~\cite{M_rigol_06,tang2013short}. We explore experimentally-accessible quantities such as the compressibility, the density of on-site pairs, and the spin-spin correlation functions as a function of temperature and interaction strength $(U/t \in [0.5,12])$. These quantities display finite-temperature signatures of the metal-insulator transition at $U_c/t \sim 6$, and capture the evolution of the magnetic ordering from 2-SL to 3-SL for $U>U_c$, which allows us to propose a ``phase diagram'' where we present regions with different characteristic behaviors of the model in the  $T-U$ space. It is important to note that these regions are not distinct phases and that boundaries correspond to crossovers characterized by short-range correlations rather than phase transitions. For example, we note the metal-insulator transition that occurs at $T=0$ and is accompanied by the presence of a long-range magnetic ordering. However at finite-$T$, the Mermin-Wagner theorem prohibits the spontaneous symmetry breaking and therefore only short-range spin correlations are present. This work provides new and experimentally accessible signatures of the metal-insulator transition and magnetic crossovers in the SU(3) FHM in the 2D square lattice.

The remainder of the paper is organized as follows. In Section~\ref{sec::Model_Methods}, we present the SU(3) FHM Hamiltonian, the observables we measure, and the details of the numerical techniques used. In Section~\ref{sec::Results}, we present the main results of the paper, and in Section~\ref{sec::Conclusions} we present our conclusions.

\section{Model and methods}\label{sec::Model_Methods} 

The Hamiltonian for the SU(3) FHM is
\begin{equation}\label{eq:Hubbard_N1}
H = -t \sum_{\langle i,j \rangle, \sigma} \left( c_{i \sigma}^\dagger c_{j \sigma}^{\phantom{\dagger}} + \mathrm{h.c.} \right) + \frac{U}{2} \sum_{i,\sigma \neq \tau} n_{i \sigma} n_{i \tau} - \mu \sum_{i,\sigma} n_{i \sigma},
\end{equation} 
where $c_{i \sigma}^\dagger$ ($c_{i \sigma}^{\phantom{\dagger}} $) is the creation (annihilation) operator for a fermion with spin flavor $\sigma = 1,2,3$ on site $i = 1,2,...,N_s$ in a 2D square lattice, $N_s$ denotes the number of lattice sites, $n_{i \sigma} = c_{i \sigma}^\dagger c_{i \sigma}^{\phantom{\dagger}}$ is the number operator for flavor $\sigma$ on site $i$, and $\mu$ is the chemical potential that controls the fermion density. We work in units where $k_B = 1$ throughout the paper.

We are interested in thermodynamic quantities and correlation functions. We focus on the density
\begin{equation}
    \expect{n}= \frac{1}{N_s} \sum_{i,\sigma} \expect{n_{i \sigma}},
\end{equation}
the isothermal compressibility
\begin{equation}
    \kappa = \frac{1}{\expect{n}^2} \frac{d \expect{n}}{d\mu},
\end{equation}
the density of on-site pairs
\begin{equation}
    \mathcal{D} = \frac{1}{N_s} \sum_i \left[ \frac{1}{2}\sum_{\sigma \neq \tau} \langle n_{i \sigma} n_{i \tau} \rangle \right],
\end{equation}
and the real space spin-spin connected correlation functions
\begin{equation}\label{eq:sr} 
    C(\vec{r}\,) = \frac{1}{N_s} \sum_{\vec{r}_0} \left[ \sum_{\sigma \neq \tau} \bigg( \langle n_{\vec{r}_0, \sigma} n_{\vec{r}_0 + \vec{r}, \sigma} \rangle - \langle n_{\vec{r}_0, \sigma} n_{\vec{r}_0 + \vec{r}, \tau} \rangle \bigg) \right],
\end{equation}
where $\vec{r}_0$ runs through all the lattice vectors in the lattice. The sign of this function is shown for the schematic 2-SL and 3-SL orders in the insets of Fig.~\ref{fig::2SL_3SL}: For the 2-SL, the sign alternates in a checkerboard pattern, while in the 3-SL the sign changes every two blocks in the Manhattan distance~\bibnote{
The Manhattan distance between two points is the sum of the absolute differences of their Cartesian coordinates.}.

The above observables provide valuable knowledge about the physics of the model: the compressibility and the density of on-site pairs are useful measures of the degree to which the system is a Mott insulator, while the spin correlation functions detect magnetic ordering, and as we will see below, capture important information about the nature of the transition. 

In practice, at each $U/t$ and $T/t$ the average density is fixed to $\expect{n} = 1.000 \pm 0.006$. In the calculation of derivatives in DQMC, we used the three-point differentiation rule~\cite{IbarraGarciaPadilla2020}.
In the case of the compressibility, the derivative is taken first and then linearly interpolated with respect to the density to obtain $\kappa$ at the target density. For NLCE, the compressibility is computed using the fluctuation dissipation theorem $\frac{d \expect{n}}{d\mu} = \frac{1}{T}\big[ \expect{n^2} - \expect{n}^2 \big]$, where $\expect{n^2} = \frac{1}{N_s^2}\sum_{i,j}\expect{n_i n_j}$ and $n_i = \sum_{\sigma}n_{i \sigma}$. Since $\expect{n}=1$ throughout the paper, we drop the prefactor $1/\expect{n}^2$ in $\kappa$ for simplicity.

Observables in thermal equilibrium are obtained using DQMC and NLCE. DQMC and NLCE are often the exact and complementary numerical methods of choice for the SU(2) FHM at the finite temperatures relevant to ultracold matter~\cite{Hart2015,Greif2013,Cheuk2016,Brown2017,Brown2018,Nichols2019,IbarraGarciaPadilla2020}. Here, we use our extension of DQMC and NLCE for SU($N$) systems~\cite{Taie2020,IbarraGarciaPadilla2021}. Generally speaking, the DQMC performs best at weak to intermediate interactions, while the NLCE performs best at strong interactions; we present results from both methods where they are viable. For the temperatures and interaction strengths where NLCE and DQMC are well converged, the methods agree well, supporting the validity of our results.

\subsection{Determinant Quantum Monte Carlo}

In this method, a path integral for the partition function is obtained by discretizing the inverse temperature $\beta$ in steps of $\Delta \tau$, i.e. $\beta = L \Delta \tau$. We use a Trotter step $\Delta \tau = 0.05/t$. We introduce three auxiliary Hubbard-Stratonovich (HS) fields to decouple each interaction term and integrate out the fermionic degrees of freedom. We then sample the HS fields stochastically. In order to obtain accurate results, we obtain DQMC data for $40-60$ different random seeds for each of the $T/t$ and $U/t$ considered. For each Monte Carlo trajectory we perform $2000$ warm-up sweeps and $8000$ sweeps for measurements. The number of global moves per sweep to mitigate possible ergodicity issues~\cite{Scalettar1991} is set to 4~\bibnote{These global moves update, at a given lattice site, all the imaginary time slices that couple two spin flavors.}. 

We analyze the systematic errors (finite-size and Trotter errors) at $T/t \leq 1$ (see Appendices \ref{App::FSS} \& \ref{App::Trotter}). To estimate finite-size effects, we perform calculations using $6 \times 6$, $8 \times 8$, and $12 \times 12$ lattices at interaction strengths above and below the quantum critical point. We find that when significant, these are inconsequential to the main conclusions of the paper. Hence, unless explicitly stated, we draw our conclusions based on results on the $6 \times 6$ lattice due to computational cost (see Sec.~\ref{sec::Results} and Appendix~\ref{App::FSS}). 
Finally, we estimate Trotter errors at large interaction strengths. These are presented in Appendix~\ref{App::Trotter}, and do not affect the main conclusions of the paper. 

\subsection{Numerical Linked Cluster Expansion}

In the NLCE, extensive properties of the lattice model in the thermodynamic limit are expressed as sums of contributions from all topologically distinct clusters, up to a certain size, that can be embedded in the lattice. Those contributions are in turn calculated exactly using full diagonalization of the model Hamiltonian on the finite clusters. See Ref.~\cite{tang2013short} for details.
We use spin flavor conservation symmetry to block diagonalize the Hamiltonian matrix of each cluster
~\cite{IbarraGarciaPadilla2021}, which allows us to perform the calculations up to seven orders in a site expansion (considering all clusters up to seven sites) 
for thermodynamic quantities,  and up to six orders for correlation functions. 

To compute correlation functions beyond nearest-neighbors, one needs to work with not only the topologically distinct clusters for a model involving only nearest-neighbor hopping, but rather all clusters that can be embedded on the lattice and are not related by translation or point-group symmetry operations. The number of the latter is significantly higher than the number of topologically distinct clusters and puts a severe constraint on time in our calculations.

The NLCE is an expansion directly in the thermodynamic limit, and its accuracy is controlled by the order (no statistical or systematic errors are present when the series is converged). We show the two highest orders and only show data for regions where the two highest orders agree within 1\%.

\section{Results}\label{sec::Results}

\begin{figure}[tbp!]
\includegraphics[width=\linewidth]{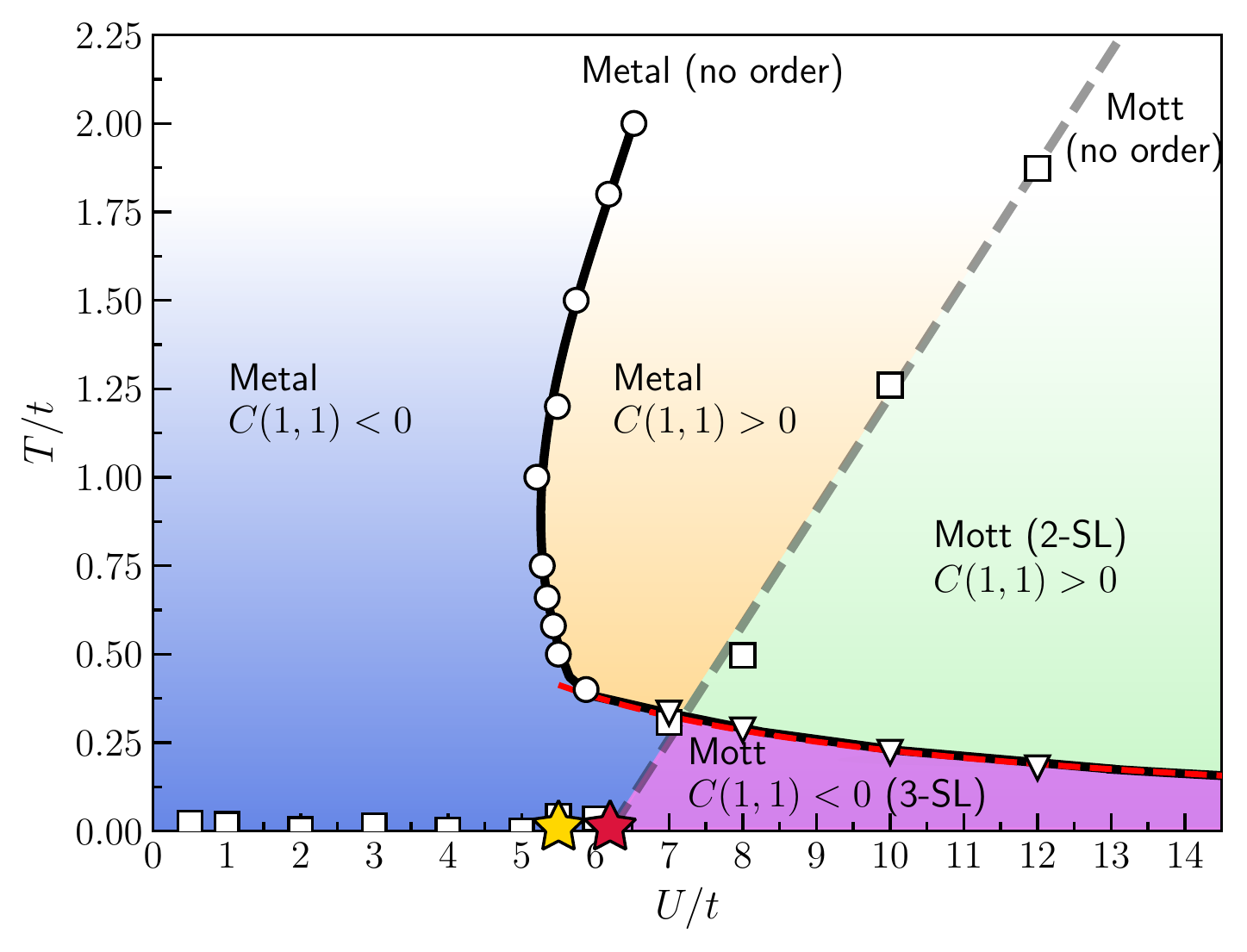}
\caption{\textbf{Regimes of the SU(3) Fermi-Hubbard model in the two-dimensional square lattice}. Squares indicate $T= \Delta$ obtained from the compressibility [see Fig.~\ref{fig::Kappa} and Eq.~\eqref{eq:fit_gap}]. Circles correspond to the zero-crossing of the next-nearest-neighbor spin correlator $C(1,1)$ [see Fig.~\ref{fig::Corr}]. Triangles correspond to the location of the dip in $dP/dT$ [see Fig.~\ref{fig::D}] (also $C(1,1)$ exhibiting a zero-crossing [see Fig.~\ref{fig::FSS_analysis}]). The black line indicates the $C(1,1)=0$ curve. The model exhibits a metal-insulator transition at $U_c/t \sim 6$ at $T=0$ (red star: this work, yellow star: CPQMC [see main text]), and crossovers at finite-$T$ indicated by the black and gray lines. 
For $U \geq U_c$ and at $T \lesssim 0.3(U-U_c)$ the system is a Mott insulator, and develops a short-range 2-SL AFM structure from a metallic phase at $T \sim J$. As the temperature is lowered, below $T/J \sim 0.57$, the system starts developing a 3-SL AFM order (the dashed red line corresponds to the $T/J=0.57$ curve). We use a qualitative color fading scheme at high temperatures to indicate the disappearance of magnetic correlations.}\label{fig::phase_diagram} 
\end{figure}

The results section is organized as follows: We first present the $T$-$U$ diagram of the SU(3) FHM which summarizes our findings. This is followed by discussions of the compressibility, which illustrates finite-temperature signatures of the metal-insulator transition. Then we analyze the spin-spin correlations, which capture the development of the short-range 2-SL AFM structure, and the change in the magnetic ordering from 2-SL to 3-SL. Finally we argue that this change in the magnetic ordering has observable effects in the density of on-site pairs. 

The main results of the manuscript are summarized in Fig.~\ref{fig::phase_diagram}. The regions shown do not represent long-range ordered phases, but instead correspond to our findings appertaining the development of short-range correlations in the model in the $T-U$ space. We will describe them here, and then describe in the rest of the paper how these conclusions were obtained. We detect finite-$T$ signatures of a metal-insulator quantum phase transition ($T=0$) consistent with a critical interaction of $U_c/t \sim 6$ (red star) by studying the compressibility. This finding is also consistent with ground state calculations using constrained path quantum Monte Carlo (CPQMC), which also predicts $U_c/t \sim 5.5$ (yellow star)~\cite{Feng2023}. From the spin-spin correlation functions we observe that for $U \gtrsim U_c$ and $T \sim J$ ($J = 4t^2/U$ is the antiferromagnetic exchange energy) a short-range 2-SL AFM structure develops and is manifest in spin correlations of about 2\% of their maximal allowed values at relatively high-temperatures of $T/t \sim 0.6$. While a fairly small value in absolute terms, correlations of this magnitude are now routinely probed in quantum gas microscope experiments~\cite{Koepsell2020,Ji2021,Hartke2020}. Furthermore, we detect in the density of on-site pairs and the spin-spin correlation function that as the temperature is lowered below $T/J \sim 0.57$, the system starts developing a 3-SL AFM order. It is worth mentioning that this change from 2-SL to 3-SL ordering is consistent with the observations in the Heisenberg limit~\cite{Romen2020} which also predict the appearance of an emerging 3-SL order, albeit at a lower temperature of $T/J \sim 0.15$. The apparent discrepancy in relevant temperatures arises from the fact that the study in the Heisenberg limit considers the evolution of the structure factor (which signals longer-range correlations), whereas here we discuss the crossover temperature, where the appropriate sign of the diagonal spin correlations appear. That is, we focus on the onset of the shortest distance correlations becoming consistent with the magnetic ordering. Naturally, this crossover temperature is expected to be higher than the temperature where the structure factor at $(\pm2\pi/3,\pm2\pi/3)$ surpasses the one at $(\pm\pi,\pm\pi)$.

\begin{figure}[tbp!]
\includegraphics[width=\linewidth]{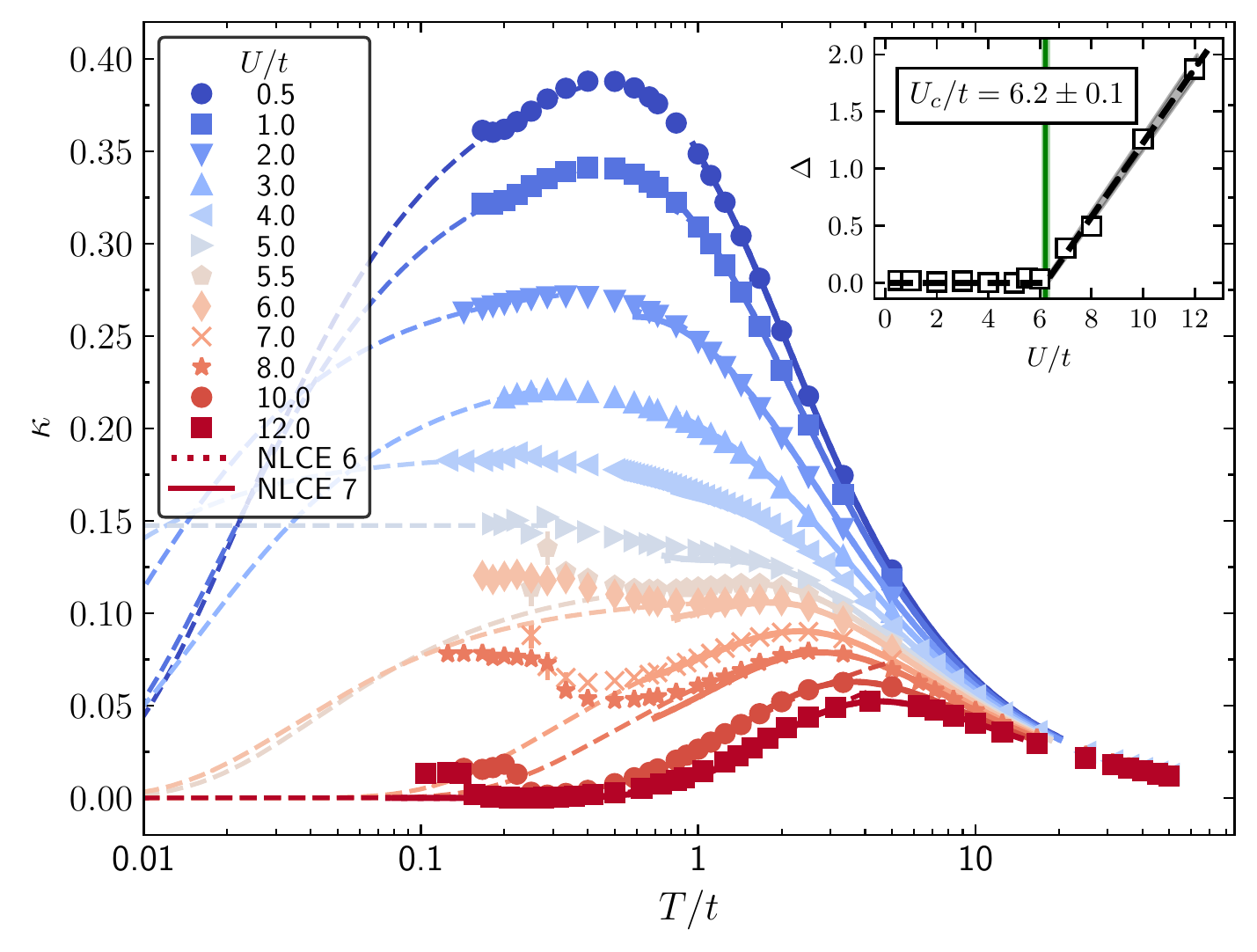}
\caption{\textbf{Compressibility as function of $T/t$ for different $U/t$ and the Mott gap as a function of $U/t$.} Markers correspond to DQMC calculations, dotted and solid lines correspond to the last two NLCE orders computed, and dashed lines are the fits to $ae^{-\Delta/T}$ to extract $\Delta(U)$. Inset: $\Delta$ as a function of $U/t$. The dashed line is obtained by fitting to Eq.~\eqref{eq:fit_gap} and indicates the location of $U_c/t=6.2\pm 0.1$ for the metal-insulator transition.}\label{fig::Kappa} 
\end{figure}

\begin{figure*}[tbp!]
\includegraphics[width=0.99\linewidth]{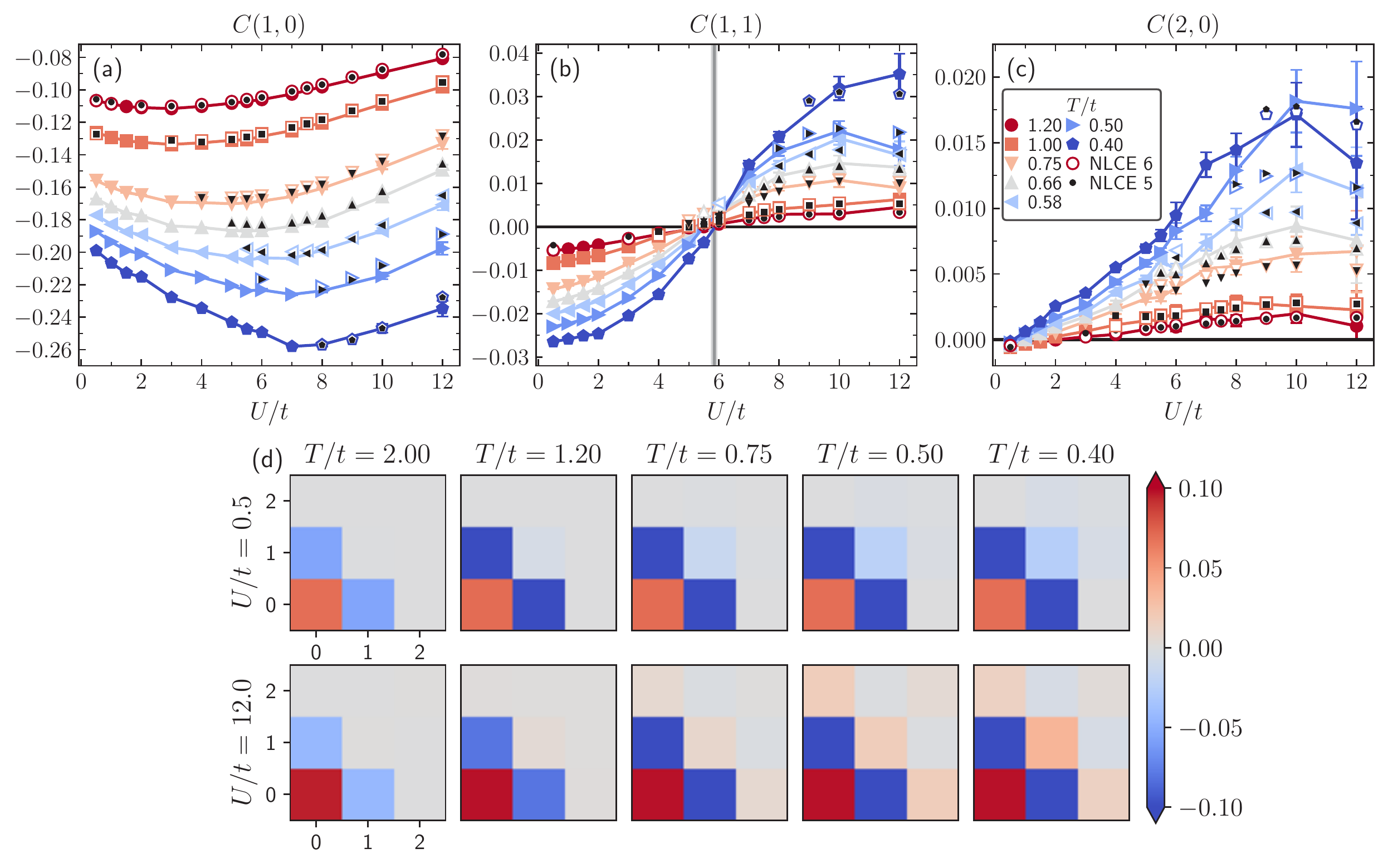}
\caption{\textbf{Spin-spin correlations}. Results are presented as a function of $U/t$ for different $T/t$ for the (a) nearest-neighbor $C(1,0)$, (b) next-nearest-neighbor $C(1,1)$, and (c) third neighbor $C(2,0)$. (d) Examples of real space spin correlations for $U/t=0.5$ and $U/t=12$ at different temperatures, which illustrate the metallic behavior of the system at small $U/t$, and the development of short-range 2-SL correlations for large $U/t$. Here the on-site correlator $C(0,0) = 2\expect{n} - 2\mathcal{D}$ is rescaled by a factor of $1/20$ so its dependence on temperature and interaction strength is more clear. In panels (a)-(c) filled color symbols correspond to DQMC, and NLCE results from order 6 (order 5) are shown as open (black) markers. Results in panel (d) correspond to DQMC calculations. In all panels DQMC results were obtained on $6 \times 6$ lattices. The circles in Fig.~\ref{fig::phase_diagram} are obtained by locating the crossing $U^*(T)$ of $C(1,1)$ vs $U/t$ at fixed $T/t$ after linear interpolation. The gray line in panel (b) is a guide to the eye of the location of $U^*(T)$ for the temperatures presented.
}\label{fig::Corr} 
\end{figure*}

The first natural quantity to look at as a measure of the Mott insulating nature of the system is the compressibility $\kappa = d\expect{n}/d\mu$, which is presented in Fig.~\ref{fig::Kappa}. Here we show $\kappa$ as a function of $T/t$ for different values of $U/t$ from DQMC (symbols) and NLCE (dotted and solid lines), which show a good agreement in their common regions of validity. For $U/t \lesssim 4$, $\kappa$ is large and the system is highly compressible, characteristic of a metal. In contrast, for $U/t \gtrsim 7$ the compressibility does not rise as strongly with increasing temperature around $T \lesssim U$ and remains small for all $T/t$. Finally, the intermediate region $4 \lesssim U/t \lesssim 7$ is the most interesting one as it illustrates a qualitatively different behavior as $T/t$ is lowered: instead of reaching a peak and then decreasing with decreasing $T$, it remains roughly flat for about a decade in temperature ($0.2 \lesssim T \lesssim 2$). It is worth noting that at these temperatures and interactions, the flatness of the compressibility also coincides with the almost temperature independent location of the $C(1,1)=0$ boundary in Fig.~\ref{fig::phase_diagram} (denoted by circles, see further discussion below).

We can distill physically salient information from this data by following the analysis performed in Refs.~\cite{Garwood2022,Kokalj2013} for the triangular lattice SU(2) Hubbard model. There, for a given $U/t$ the data is fitted using $\kappa = ae^{-\Delta/T}$ (where $\Delta$ is the charge gap) in the temperature window where $\kappa$ exhibits a non-negative slope. These fits are indicated as dashed lines in Fig.~\ref{fig::Kappa} and allow us to extract $\Delta(U)$ which we then plot as a function of $U/t$ (see inset in Fig.~\ref{fig::Kappa}). These results show that the system is metallic below $U/t \sim 6$ and insulating above it, with a linear dependence in $U/t$ as expected. To extract the quantum critical point from this technique we perform the following fit
\begin{equation}\label{eq:fit_gap}
\Delta =
\begin{cases}
0 & \text{if $U<U_c$} \\
b(U-U_c) & \text{if $U>U_c$} \\
\end{cases},
\end{equation}
for which we obtain the parameters $b= 0.32 \pm 0.01$, and $U_c = 6.2 \pm  0.1$~\bibnote{We tested the method for the half-filled SU(2) FHM in the honeycomb lattice where $U_c/t \sim 3.8$ ($3.87$ from Ref.~\cite{sorella2012} and $3.78$ from Ref.~\cite{Assaad2013}). This method is consistent with the results, predicting $U_c/t = 3.9 \pm 0.2$. Although
these results partially support the validity of the analysis
in the present study, this technique isn't perfectly reliable ---for example, it gives a finite $U_c$ in the SU(2) square lattice case--- and so, confirmation with other techniques is advised (such as CPQMC as in this case).}.

\begin{figure}[tbp!]
\includegraphics[width=\linewidth]{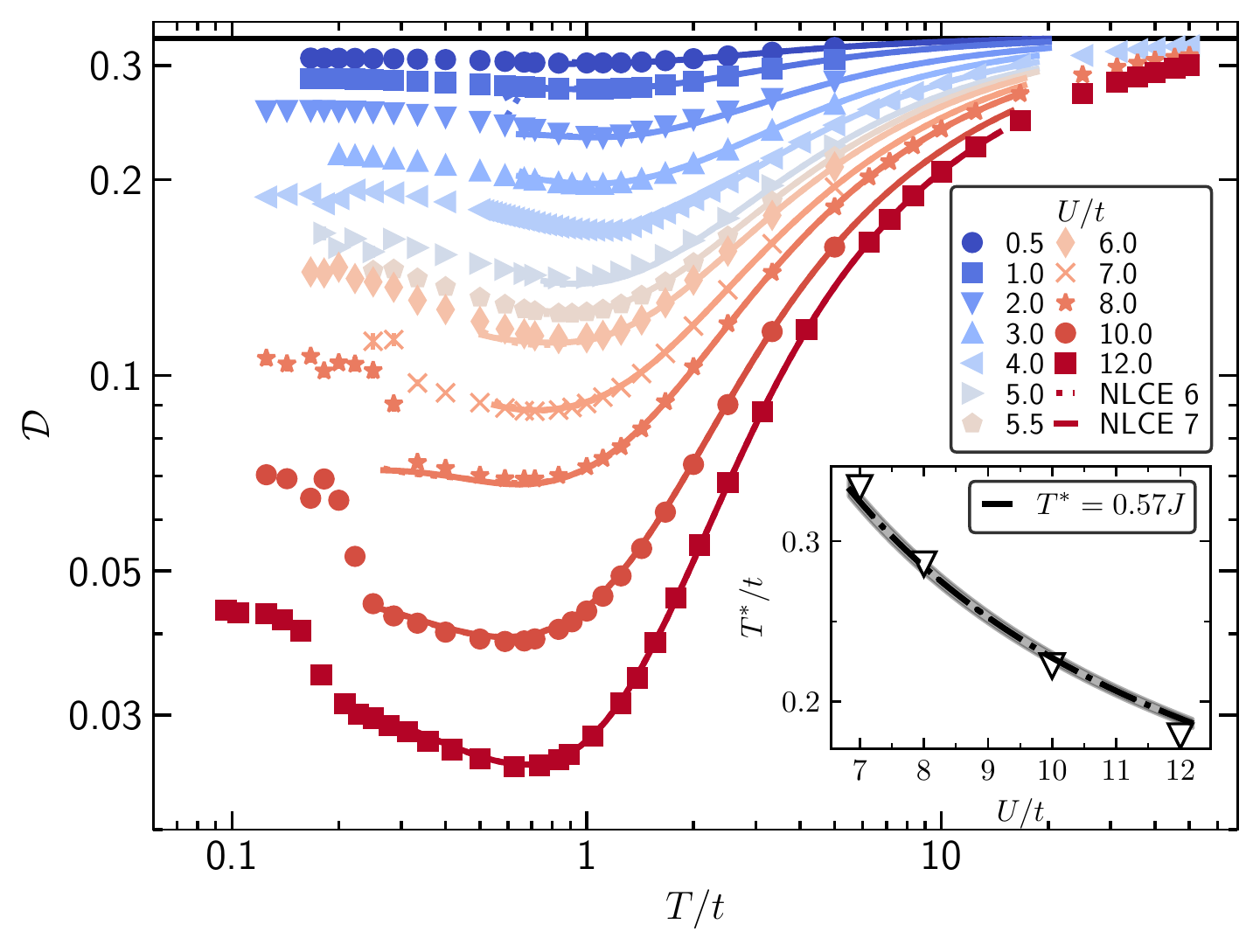}
\caption{\textbf{Density of on-site pairs}. DQMC (markers) and NLCE (lines) results are presented as a function of $T/t$ for different $U/t$ for the density of on-site pairs $\mathcal{D}$. The horizontal black line at $\mathcal{D} = 1/3$ corresponds to the $U=0$ limit. The inset presents the behavior of $T^*/t$ as a function of $U/t$, where $T^*/t$ are obtained by locating the minima of  $d(U \mathcal{D})/dT$ for $U/t >5$. The data are fitted to $T^* = (0.57 \pm 0.01) J$.
}\label{fig::D} 
\end{figure}

Finally, it is worth mentioning that the range of temperatures included in the fit over $T/t$ at fixed $U/t$ omits the temperatures below the upturn in $\kappa$ at the lowest $T/t$ for $U/t >6$, which occurs at the same temperature scale where $\mathcal{D}$ exhibits a sharp upturn and is related to a change in the magnetic ordering (see Fig.~\ref{fig::D} and its discussion for more details).

At $T=0$ the gap opening may coincide with the magnetic ordering transition,  however at finite temperatures these may behave differently. To investigate this, Fig.~\ref{fig::Corr} presents the spin-spin correlations as a function of $U/t$ for different $T/t$. Fig.~\ref{fig::Corr}(a) shows the nearest-neighbor spin correlation $C(1,0)$, which highlights the agreement between DQMC and NLCE. We first note that $C(1,0)<0$ for all values of $U/t$. As temperature is lowered, the AFM nearest-neighbor correlations increase for all $U/t$, being strongest around $U/t \sim 7$ for the lowest temperatures. At the lowest temperatures, the shape of $C(1,0)$ with $U/t$ can be understood as follows: Starting from the non-interacting limit, as one increases $U/t$, local moment formation will lead to the development of stronger AFM nearest-neighbor correlations; however when increasing $U/t$ in the strong-coupling region, the relevant energy scale becomes the superexchange energy, $J=4t^2/U$, and thus the correlations decrease for larger $U/t$ at fixed $T/t$.

The next-nearest-neighbor spin correlator $C(1,1)$ exhibits a distinctive $U/t$ dependence [Fig.~\ref{fig::Corr}(b)], where again results from the two methods agree. At temperatures above $T/t \geq 2$, $C(1,1)$ vanishes for all $U/t$ considered and no order is expected [see, for example, Fig.~\ref{fig::Corr}(d)]. As the temperature is lowered, the correlations increase in magnitude and at each temperature a zero-crossing occurs at $U^*(T)$ (indicated with circles in Fig.~\ref{fig::phase_diagram}), i.e. $C(1,1)<0$ for $U < U^*(T)$ and changes sign for $U > U^*(T)$. This phenomena is reminiscent of the doped SU(2) FHM on square lattices where $C(1,1)$ changes sign upon doping at a fixed $T/t$ and $U/t$~\cite{Cheuk2016}. In the SU(2) case $C(1,1)<0$ indicates metallic behavior as it reflects the effects of Pauli blocking, a signature of non-interacting fermions. In our case, the change in sign of $C(1,1)$ reflects a combination of Pauli blocking (dominant at $U<U^*$) and interaction-driven superexchange physics ($U>U^*$).

Figure~\ref{fig::Corr}(c) illustrates the third neighbor spin correlation function $C(2,0)$. We find $C(2,0)>0$ for most values of $U/t$ shown (generally stronger for larger values of $U/t$), and it increases in magnitude as the temperature is lowered. The positive sign of the correlator is in agreement with the sign of the spin correlator in the 2-SL ordering [see inset in Fig.~\ref{fig::2SL_3SL}(a)]. We attribute the disagreement of DQMC results at $T/t\le 0.5$ in the strong-coupling region with the converged NLCE results to systematic errors in the DQMC, especially the finite-size error for this longer range correlation function. Finally, Fig.~\ref{fig::Corr}(d) presents the real-space spin-spin correlation functions for $U/t=0.5$ and $12$ for different $T/t$. As the temperature is lowered, the system develops stronger nearest-neighbor AFM correlations for both $U/t$, and for $T/t =1.2$ the sign change of $C(1,1)$ is already evident. As the temperature is lowered even further, the system at weak $U/t$ remains metallic, but at strong $U/t$ displays a short-range 2-SL AFM order in agreement with the Heisenberg limit predictions~\cite{Romen2020}. It is important to point out that the study of the nature of the underlying 2-SL structure is an interesting question that requires further exploration, and we expect future studies to be able to resolve it by analyzing correlation functions of the order parameters derived from the Gell-Mann matrices (see Ref.~\cite{Sotnikov2015} for a discussion).

\begin{figure}[hbtp!]
\includegraphics[width=\linewidth]{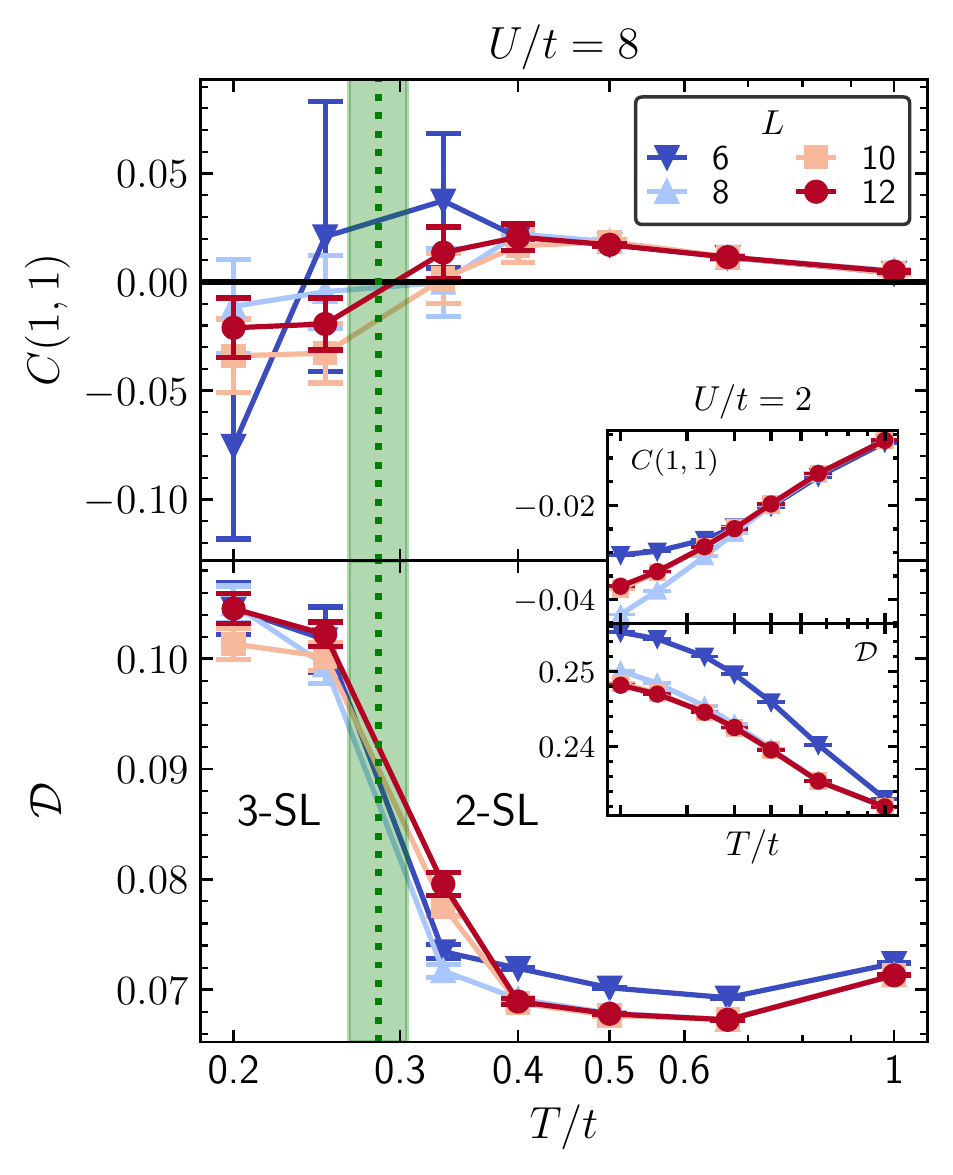}
\caption{\textbf{Finite-size analysis}. $C(1,1)$ and $\mathcal{D}$ as a function of $T/t$ are presented for $U/t=8$ ($U/t=2$ in the inset) at system sizes: $6\times6$, $8\times8$, $10\times10$, and $12\times12$. Dotted vertical line corresponds to $T = 0.57 J$, and the shaded region is half the spacing between the middle point of the two consecutive markers. Horizontal scale of the inset is the same as the main figure.}\label{fig::FSS_analysis} 
\end{figure}

Systematic calculations of $C(1,1)$ for different system sizes further support the claim of the change in sign observed in Fig.~\ref{fig::Corr}(b) for $T/t \geq 0.4$. This is illustrated in the upper panels of Fig.~\ref{fig::FSS_analysis} where we show results for $U/t=8$ (main panel) and $U/t=2$ (inset) as a function of $T/t$ for different system sizes $L \times L$. As the temperature is lowered below $T/t < 0.4$ we observe that for $U/t=8$ finite-size errors are severe for $6\times6$ lattice. However,  results for $L\geq8$ point to the same behavior: they all display a zero-crossing in the temperature window $ 0.25 < T/t <0.33$. Moreover, results for the two largest system sizes, $10\times10$ and $12\times12$, agree with each other within the error bars and make clear that $C(1,1)$ is negative at the lowest two temperatures and positive for the rest. For these system sizes, we compute the correlation function at the three lowest temperatures for $U/t =8$ using 160 independent runs to improve the statistics. We identify this change in sign at lower temperatures as the temperature scale at which the change in ordering from 2-SL to 3-SL occurs [see insets in Fig.~\ref{fig::2SL_3SL} which illustrate the sign of the spin correlation function for both orderings]. In contrast, for $T/t < 0.4$ and $U/t=2$, despite the presence of finite-size errors in $6\times6$ and $8\times8$ lattices, results for the two largest system sizes are consistent with each other and remain negative.

This change in sign in $C(1,1)$ as a function of temperature coincides with the low-temperature rise in the density of on-site pairs $\mathcal{D}$ for $U/t=8$ as evidenced in the lower panel of Fig.~\ref{fig::FSS_analysis}. For $U/t=8$, the finite-size error in $\mathcal D$ is minimal at all temperatures considered in $6\times 6$ lattices, and the location of the pronounced upturn remains consistent across all system sizes. For comparison, for $U/t=2$ there is a finite-size error in $\mathcal{D}$, but it does not change the smooth temperature dependence of $\mathcal{D}$ as demonstrated by calculations on $12\times 12$ lattices, which agree with the $6\times 6$ results (see lower inset in Fig.~\ref{fig::FSS_analysis}).

These findings allow us to conclude that the change in the magnetic ordering from 2-SL to 3-SL, as evidenced by the change in sign of $C(1,1)$ for $U>U_c$, has observable effects in the density of on-site pairs: the temperature scale at which the upturn in $\mathcal{D}$ occurs coincides with that of the zero-crossing. This connection between $\mathcal{D}$ and the change in magnetic ordering motivates our next analysis. 

Unlike the correlation function $C(1,1)$, $\mathcal{D}$ exhibits minimal finite-size errors. The low-temperature upturn can also be captured using less statistics, and hence requires less computational resources, than that needed to resolve the 2-SL to 3-SL magnetic crossover using $C(1,1)$. In Fig.~\ref{fig::D}, we show the density of on-site pairs $\mathcal{D}$ as a function of $T/t$ for different values of $U/t$, which displays several familiar trends as well as the novel low-temperature feature. As expected, larger $U/t$ lead to smaller overall $\mathcal{D}$, and as the temperature initially decreases, $\mathcal{D}$ is suppressed from its high-temperature value around $T \sim U$. Following the development of antiferromagnetic correlations in the system below a temperature of the order of $J$,  $\mathcal{D}$ slowly rises again upon further lowering the temperature. That is because the probability of finding a particle with a different flavor at, and therefore virtual hoppings to, neighboring sites increases.

In addition to the above increase, we observe the remarkably sharp upturn in $\mathcal{D}$ at even lower temperatures for $U>U_c$ only. As previously discussed, we attribute this behavior to a magnetic reordering of spins from the 2-SL to the 3-SL ordered phases. The inset in Fig.~\ref{fig::D} presents the location $T^*/t$ vs $U/t$, where $T^*/t$ is defined as the temperature at which the minima of $d(P =U \mathcal{D})/dT$ is located, namely, the temperature where the upturn of $\mathcal{D}$ upon lowering $T$ is the largest (see Fig.~\ref{fig::dPdT_app} in the Appendix for the plot of $dP/dT$ vs $U/t$) and corresponds to the change in magnetic ordering. We then fit the points to $T^* = a J$, with $a = 0.57 \pm 0.01$. It is important to point out that this phenomenon survives the extrapolation $\Delta \tau \to 0$, as illustrated in Fig.~\ref{fig::Trotter_analysis} of Appendix~\ref{App::Trotter}, and finite-size scaling as shown in Fig.~\ref{fig::FSS_analysis}. On the other hand, the NLCE results do not capture this phenomenon as the lowest convergence temperatures are typically above the onset of this sharp upturn. 

\section{Discussion and outlook}\label{sec::Conclusions}

We have explored the behavior of the SU(3) FHM in the square lattice at $1/3$-filling as a function of $U/t$ and $T/t$ using complementary DQMC and NLCE methods. We computed the compressibility, the density of on-site pairs, and the spin-spin correlation functions, which are experimentally accessible with ultracold AEAs in OLs. These quantities express finite-temperatures signatures of a quantum metal-insulator transition, which we locate at $U_c/t \sim 6$, and capture the temperature scale where the magnetic ordering evolves from a short-range 2-SL to a 3-SL for $U>U_c$ at $T/J \sim 0.57$. 

These results imply the existence of different regimes of the model in the $T-U$ plane, which are accessible in current experiments with AEAs in a 2D OL. First, the compressibility can be computed by measuring the equation of state $\expect{n(\mu,U,T)}$ and taking the derivative of $\expect{n}$ with respect to $\mu$~\cite{Hofrichter2016}. Furthermore, local access to the density in SU($N$) systems~\cite{Pasqualetti2023} allows experiments to probe the density fluctuations, and therefore extract the temperature, since these are related to the compressibility via the fluctuation dissipation theorem~\cite{Qi2011,Hartke2020}. Second, the density of on-site pairs can be measured using a photoassociation technique~\cite{Ozawa2018,Taie2020,Hofrichter2016}, which can be used to locate $T^*$. Finally, $C(1,1)$ can be directly measured with an AEA quantum gas microscope. The low temperatures of some features seen here present some challenge, but combining the temperatures already achieved in Ref.~\cite{Taie2020} with local imaging should allow experiments to observe most of the phenomena presented in Fig.~\ref{fig::phase_diagram}. 

In fact, the key quantities to see the physics described in this paper can be measured without a full quantum gas microscope, as long as local resolution is good enough to measure the density and its fluctuations, as has been demonstrated for SU($N$) AEA systems very recently~\cite{Pasqualetti2023}. We discuss a protocol for this here: the idea is to relate spin fluctuations in a small region $W$ set by the imaging resolution to number fluctuations using a shelving scheme, and then relate  the spin fluctuations to correlations within the region $W$.

The idea is to remove or shelve spin components, e.g. by using the 578nm narrow clock transition in  $^{173}$Yb, and measure the density fluctuations of the remaining ground state atoms.  First note that the total particle number measured in a region $W$ is $n_W = \sum_{j \in W} n_j$ (where $n_j = \sum_\sigma n_{j\sigma}$), and the averaged fluctuations of this quantity, $\mathcal C$ are
\begin{align}
\mathcal C &= \braket{n_W^2}-\braket{n_W}^2 \\
 &=   \braket{  \big(\sum_{j \in W} n_j\big)^2}- \braket{  \sum_{j \in W} n_j}^2\\
 	&= \sum_{i,j\in W} \left(\braket{n_i n_j} -\braket{n_i}\braket{n_j}\right) \\
  &= \sum_{i,j\in W;\sigma \tau} \bigg[\braket{n_{i\sigma} n_{j\tau}} -\braket{n_{i\sigma}}\braket{n_{j\tau}}\bigg] \label{eq:density-corr-fluc}
\end{align}
Hence we see the number fluctuations in $W$ are related in a simple way to the correlations. 

Eq.~\eqref{eq:density-corr-fluc} can be used to relate density fluctuations to spin correlations by shelving $M$ of the $N$ components (any $0<M < N$ suffices) to states that are not imaged. Using $\braket{\cdots}$ to represent expectations in the pre-shelved state (which we want to know) and $\braket{\cdots}_{\text{shelved}}$ for expectations in the shelved state, we obtain for the density fluctuations after shelving
\begin{align}
\mathcal C^{(M)}  &=\sum_{i,j\in W;\sigma \tau} \bigg[\braket{n_{i\sigma} n_{j\tau}} -\braket{n_{i\sigma}}\braket{n_{j\tau}}\bigg]_{\text{shelved}}  \\
&= \sum_{i,j\in W;\sigma \tau \in \text{unshelved}} \bigg[\braket{n_{i\sigma} n_{j\tau}} -\braket{n_{i\sigma}}\braket{n_{j\tau}}\bigg]
\end{align}
The second line follows because if spin component $\nu$ is shelved, any terms involving $n_{i\nu}$ vanish while unshelved states are unaltered. Using the permutation symmetry of the spin components,
\begin{align}
 \mathcal C  &=  N 
  \sum_{i,j\in W} \bigg[\braket{n_{i\sigma} n_{j\sigma}} + (N-1)\braket{n_{i\sigma} n_{j\tau \neq \sigma}} \nonumber \\
  & \qquad \qquad \quad  -N \braket{n_{i\sigma}}^2\bigg] \\
 \mathcal C^{(M)}  &= M   
 \sum_{i,j\in W} \bigg[\braket{n_{i\sigma} n_{j\sigma}} + (M-1)\braket{n_{i\sigma} n_{j\tau \neq \sigma}} \nonumber \\
 & \qquad \qquad \quad  -M \braket{n_{i\sigma}}^2\bigg]
\end{align}
And the sum of the spin-spin correlation function [defined in Eq.~\eqref{eq:sr}] in region $W$ is therefore:
\begin{equation}\label{eq::protocol}
    \tilde{C}_s = \frac{N-1}{M(N-M)}\bigg[N^2 \mathcal{C}^{(M)} - M^2\mathcal{C} \bigg].
\end{equation}
If $T \gtrsim J$, where we can assume only spin correlations up to next-nearest-neighbors contribute, Eq.~\eqref{eq::protocol} then gives a combination of $C(0,0)$, $C(1,0)$ and $C(1,1)$. $C(0,0)$ is related to $\mathcal{D}$ via $C(0,0) = (N-1)\expect{n} - 2\mathcal{D}$, so it can be measured independently as previously discussed, while $C(1,0)$ can be measured independently using the singlet-triplet oscillation method~\cite{Ozawa2018,Taie2020}. With these measurements and Eq.~\eqref{eq::protocol} $C(1,1)$ can be resolved.

A major goal in strongly correlated matter is to understand the consequences of doping Mott insulators and magnetically ordered phases.
Our results provide a useful starting point for the exploration of the doped SU(3) FHM at finite-temperature. This has been an active area in the past few years for SU(2) ultracold FHMs. One example of progress is the development and testing of geometric-string theory~\cite{Chiu2019,Koepsell2020,Grusdt2018,Grusdt2019,Bohrdt2020,Bohrdt2021a,Bohrdt2021,Bohrdt2022,Grusdt2023}, which draws a connection between the strongly correlated quantum states at finite doping and the AFM parent state at $1/2$-filling. Ref.~\cite{Chiu2019} found that for $N=2$ the geometric string patterns are observable at $T \lesssim J$ and upon $\lesssim 10\%$ doping. These temperatures have been attained with AEAs in OLs and therefore the exploration of the doped SU(3) FHM is a timely question. Our results at $1/3$-filling exhibit interesting magnetic crossovers around $T^* =0.57J$, and we expect this to drive qualitative changes in the  string length and anisotropy across this crossover.

\begin{acknowledgments}
Authors would like to thank useful discussions with Yoshiro Takahashi, Shintaro Taie, Henning Schl\"omer, and Sohail Dasgupta. EIGP, RTS, and EK are supported by the grant DE-SC-0022311, funded by the U.S. Department of Energy, Office of Science. We acknowledge support from the Robert A. Welch Foundation (C-1872), the National Science Foundation (PHY-1848304), and the W. F. Keck  Foundation (Grant No. 995764). Computing resources were supported in part by the Big-Data Private-Cloud Research Cyberinfrastructure MRI-award funded by NSF under grant CNS-1338099 and by Rice University's Center for Research Computing (CRC),
as well as the Spartan high-performance computing facility at San José State University supported by the NSF under Grant No. OAC-1626645. KRAH's contribution benefited from discussions at the Aspen Center for Physics, supported by the National Science Foundation grant PHY-1066293, and the KITP, which was supported in part by the National Science Foundation under Grant No. NSF PHY-1748958. 
\end{acknowledgments}

\appendix

\section{Interaction energy derivative}\label{App::dPdT}

\begin{figure}[hbtp!]
\includegraphics[width=\linewidth]{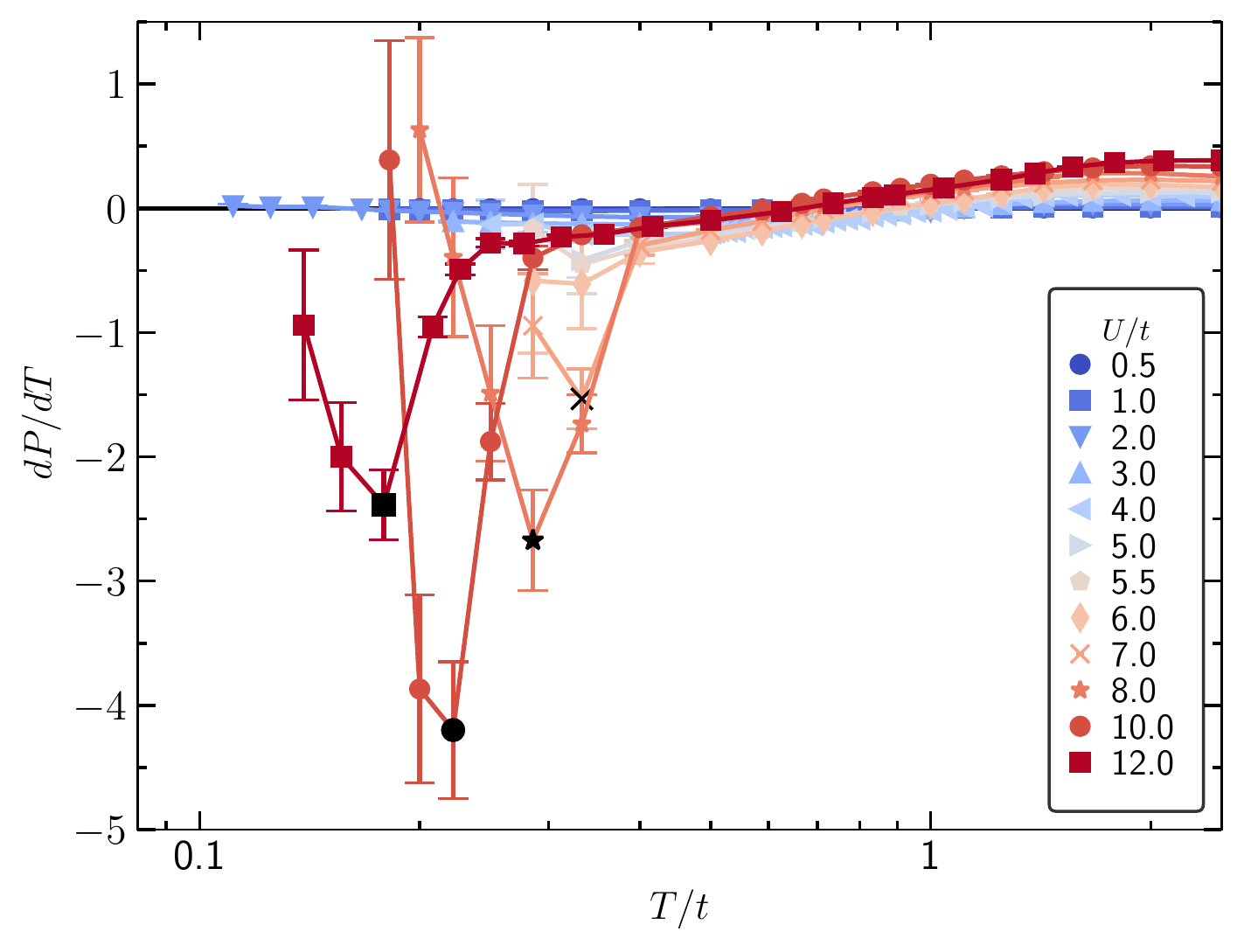}
\caption{\textbf{Derivative of the interaction energy $P = U \mathcal{D}$ with respect to temperature $d(U \mathcal{D})/dT$ vs $T/t$}. Results are presented for different $U/t$. Black markers correspond to $T^*/t$ which are obtained by locating the minima of $dP/dT$ for $U/t >5$. Solid markers are DQMC data and lines connect the dots to guide the eye.}\label{fig::dPdT_app} 
\end{figure}

Fig.~\ref{fig::dPdT_app} shows the derivative of the interaction energy $P = U \mathcal{D}$ with respect to temperature, $dP/dT$. Here, we observe that only for $U/t > 6$ does $dP/dT$ exhibit a clear sharp dip when temperature increases at low temperatures. For those interaction strengths, we define $T^*/t$ as the temperature at which the minima of $dP/dT$ is located and we indicate them with black markers in this figure. They are shown as a function of $U/t$ in the inset of Fig.~\ref{fig::D}. 

\section{Finite-size effects}\label{App::FSS}

\begin{figure}[htbp!]
\includegraphics[width=\linewidth]{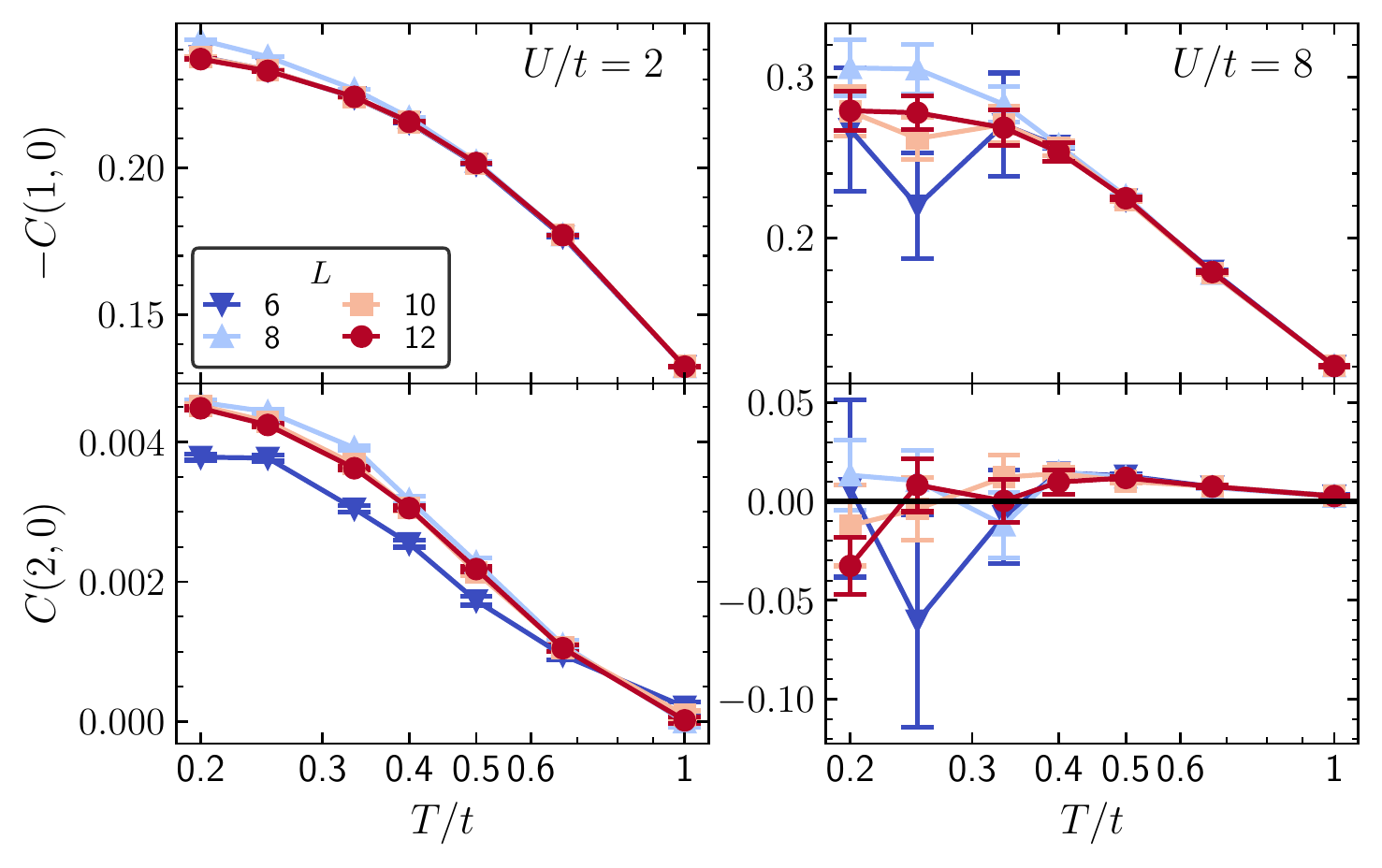}
\caption{\textbf{Finite-size analysis}. Spin correlators as a function of $T/t$ are presented for $U/t=2$ and $U/t=8$ at system sizes: $6\times6$, $8\times8$, $10\times10$, and $12\times12$. 
}\label{fig::FSS_analysis_app} 
\end{figure}

In Fig.~\ref{fig::FSS_analysis} we presented $\mathcal{D}$ and $C(1,1)$ for $L\times L$ systems with $L=6,8,10,12$. In Fig.~\ref{fig::FSS_analysis_app} we present $C(1,0)$ and $C(2,0)$ for completeness. For $U/t=2$, $C(1,0)<0$ and is well converged at all temperatures already for $L=6$. For $C(2,0)$, finite-size effects are minimal for $8\times8$ lattices and the value of the correlation always remains positive but very small. On the other hand for $U/t=8$, both spin correlations are well converged for $T>0.4$, but finite-size errors are present for $6\times6$ lattices at $T/t<0.4$. At this interaction, $C(1,0)<0$ for all system sizes and is well converged for $L=10$ and $12$. Finally, $C(2,0)$ requires further statistics to be resolved at the lowest temperatures, which are inaccessible to our current numerical capabilities.

\section{Analysis of Trotter error}\label{App::Trotter}

\begin{figure}[htbp!]
\includegraphics[width=\linewidth]{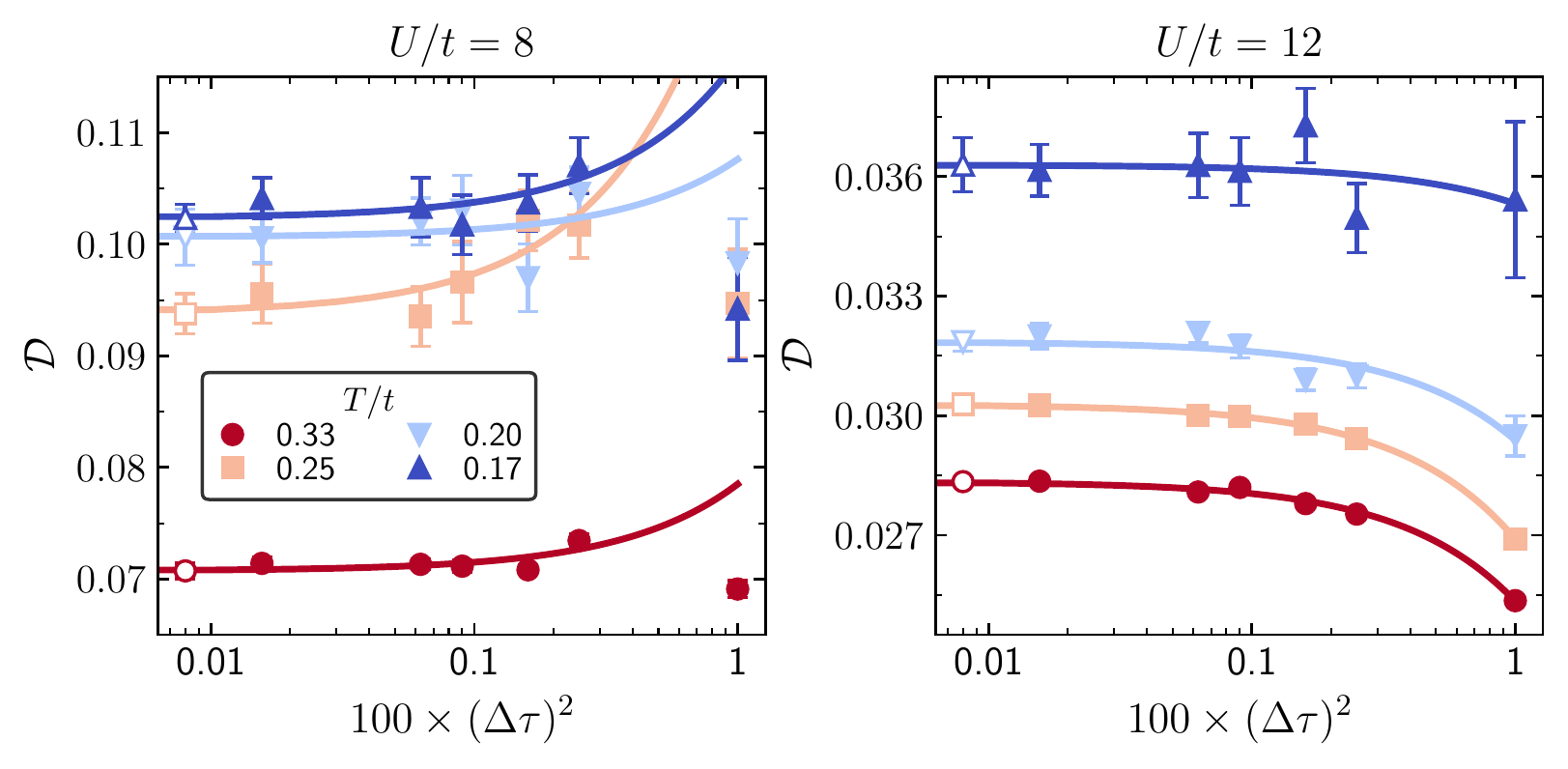}
\caption{\textbf{Density of on-site pairs as a function of $(\Delta \tau)^2$}. Results are presented for different $T/t$ for $U/t=8$ and $U/t=12$. Solid markers indicate DQMC results, solid lines correspond to linear fits to the data, and open markers are the $\Delta \tau \to 0$ extrapolation obtained from the fit.}\label{fig::Trotter_analysis} 
\end{figure}

The density of on-site pairs $\mathcal{D}$ has historically been found to be one of the most sensitive quantities to the systematic Trotter error~\cite{Chang2013}. In order to support the claims derived from the low-temperature behavior observed at large $U/t$ we analyze the Trotter error by comparing the results obtained for $\mathcal{D}$ with $\Delta \tau  = 0.0125/t$, $0.025/t$, $0.03/t$, $0.04/t$, $0.05/t$, and $0.1/t$ at $T/t= 0.17, 0.2, 0.25$, and $0.33$ for $U/t =8$ and $12$. These are presented in Fig.~\ref{fig::Trotter_analysis}. In all cases, the $\Delta \tau \to 0$ extrapolation leads to a correction no larger than $\sim 5\%$, much smaller than the low-temperature rise of $\mathcal{D}$ for $U>U_c$.

\bibliography{SU3_magnetism.bib}

\begin{thebibliography}{70}
\bibitem [{1()}]{1}%
  \BibitemOpen
  \href@noop {} {}\bibinfo {note} {The Manhattan distance between two points is
  the sum of the absolute differences of their Cartesian
  coordinates.}\BibitemShut {Stop}%
\bibitem [{2()}]{2}%
  \BibitemOpen
  \href@noop {} {}\bibinfo {note} {These global moves update, at a given
  lattice site, all the imaginary time slices that couple two spin
  flavors.}\BibitemShut {Stop}%
\end{thebibliography}

\end{document}